\documentclass[a4paper,english,aps,manuscript]{revtex4}
\usepackage[T1]{fontenc}
\usepackage[latin1]{inputenc}
\usepackage{graphicx}
\usepackage{amssymb}

\makeatletter

\usepackage{babel}
\makeatother
\begin{document}

\title{Macroscopic electrostatic potentials and interactions in self-assembled
molecular bilayers: the case of Newton black films.}

\author{Z. Gamba}

\affiliation{Department of Physics - CAC, Comisión Nacional de Energía Atómica,
Av. Libertador 8250, (1429) Buenos Aires, Argentina. }

\email{gamba@tandar.cnea.gov.ar. URL: http://www.tandar.cnea.gov.ar/~gamba}

\date{\today}

\begin{abstract}
We propose a very simple but 'realistic' model of amphiphilic bilayers,
simple enough to be able to include a large number of molecules in
the sample, but nevertheless detailed enough to include molecular
charge distributions, flexible amphiphilic molecules and a reliable
model of water. All these parameters are essential in a nanoscopic
scale study of intermolecular and long range electrostatic interactions.
We also propose a novel, simple and more accurate macroscopic electrostatic
field for model bilayers. This model goes beyond the total dipole
moment of the sample, which on a time average is zero for this type
of symmetrical samples, i. e., it includes higher order moments of
this macroscopic electric field. We show that by representing it with
a superposition of gaussians it can be \emph{analytically} integrated,
and therefore its calculation is easily implemented in a MD simulation
(even in simulations of non-symmetrical bi- or multi-layers). In this
paper we test our model by molecular dynamics simulations of Newton
black films.
\end{abstract}
\maketitle

\section{Introduction}

The structure and dynamics of amphiphilic bilayers plays a key role
in numerous problems of interest in physical-chemistry \cite{chandler},
biology \cite{mike-polymer}, pharmacology \cite{mike-anesthetic}
and biotechnology \cite{nanotech3,biotech-1}. Amphiphilic molecules
consist of a non-polar hydrophobic flexible chain of the $(CH_{2})_{n}$
type, the 'tail', plus a hydrophilic section, a strongly polar 'head'
group. In aqueous solutions the polar 'head' strongly interacts with
water and shields the hydrophobic tails from the water, therefore
these molecules tend to nucleate in miscelles or bilayers, depending
on concentration \cite{chandler}. For example, a simple model of
a biological membrane consists of a bilayer of amphiphilic molecules,
with their polar heads oriented to the outside of the bilayer and
strongly interacting with the surrounding water. The opposite model
of bilayer, with the water in the middle and head groups pointing
to the inside, also exists in nature and are the soap bubbles films,
or Newton black (NB) films, as they are called in their thinnest states.
Most studies have been performed on biological membrane models and
huge advances in the intermolecular potential models (at atomistic
or coarse grained level) needed for their numerical simulations as
well as in the behavior of proteins, anesthetics molecules , pores,
etc., inserted in these bilayers have been achieved . \\

Here we will limit our study to the case of soap bubbles films. While
on one hand the molecular interactions and macroscopic electrostatic
field exhibit the same physical problems to solve as that of the biological
membranes, on the other hand the amount of water is limited by the
width of the NB film and therefore the numerical simulations are considerable
less expensive. \\

Furthermore, the study of foam films is not only interesting \emph{per
se}, but also has many technological applications, in particular their
use in biothechnological applications is a very active field. For
example, a protein can be unfolded when inserted in these films, allowing
a small angle X ray measurement of its structure. In refs. \cite{soap.bub.protein,soap.bub.proteins2}
the structure of the film is studied as a function of the concentration
of inserted proteins; these measurements give information on the protein-lipids
interactions, in much simpler experimental arrays than biological
membranes. At the mesoscopic scale, the diffusion of bacteria in this
quasi-two-dimensional liquid environment has been measured \cite{soap.bub.bacteria}. 

From the point of view of applied physics there is also incresing
interest in the grow of inorganic nano-patterned films with the help
of organic bilayers \cite{nanotech}.

Furthermore, recent studies have revived the interest in the basic
physics involved in these films. The self-assembly of amphiphilics
can be experimentally measured by studying the properties of soap
films, with the advantage of a fine control in the range of low contents
and small concentration changes of amphiphilics \cite{soap-assembly}.
Also confined water is known to have a very different behaviour of
bulk water, its behavior at criogenic temperatures and its glass transition
temperature were measured by anelastic spectroscopy \cite{confined.w}
in a stack of biological membranes, but not yet in foam films. In
ref. \cite{soap.bub.stability} the stability of soap films under
different applied capillary pressures has been meassured, and in ref.
\cite{soap.bub-electr.fluct} a mean field theory is developed to
describe the rôle of electrostatic fluctuations in their stability.
\\

Several MD simulations of NB films have been performed since some
time ago. In Ref. \cite{zg} a series of all-atom MD simulations of
NB films of sodium-dodecyl-sulfate SDS {[}Na$^{\textrm{+}}$CH$_{\textrm{3}}$(CH$_{\textrm{2}}$)$_{\textrm{11}}$(OSO$_{\textrm{3}}$)$^{\textrm{-}}$]
amphiphilic molecules and different amounts of water was presented,
and an insight of the experimental electron density profile given
by X-ray measurements was obtained \cite{soap.bub.thick0}. In Ref.
\cite{soap-md-faraudo} these simulations were extended to larger
samples and longer times (about nanosec.), a fact that is important
to accurately measure the diffusion constants of the lipids in bilayers,
even when similar structural and internal dynamical properties can
be obtained at shorter times \cite{mem-size-diff}. The effect of
several surfactants on the disjoining pressure \cite{goddard-06}
and the anomalous behavior of water \cite{soap.bub.dielec} were also
investigated \emph{via} MD simulations. The last reference clearly
pointed out that the study of electrostatic forces between charged
interfaces in aqueous media is still an open and very important field
of theoretical research. \\

Here we are interested in the search of a very simple but 'realistic'
model of the Newton black films, simple enough to be able to include
a large number of molecules in the sample, but nevertheless detailed
enough to include molecular charge distributions, flexible amphiphilic
molecules and a reliable model of water. All these parameters are
essential in a nanoscopic scale study of intermolecular and long range
electrostatic interactions. Such simple amphiphilic bilayer model
will be useful to obtain reliable information on the effect of the
{}``external parameters'' (like surface tension, external pressure
and temperature and external fields) on physical properties of the
bilayer, as well as to address problems like the diffusion and/or
nucleation of guest molecules of technological, pharmaceutical or
ambiental relevance within amphiphilic bilayers. \\

A problem to solve in the simulation of these highly charged bilayers
is the method to use in the calculation of electrostatic interactions,
due to their long range and quasi-bidimensional 2D nature. The Ewald
method is assumed to be the accurate one for electrostatic interactions
\cite{tildesley} in three dimensional 3D samples. Refs. \cite{solventes,cpc96,finite-crystal-qq}
review the 3D Ewald sums as well as several possible implementations,
in those papers 3D periodic boundary conditions are used to minimize
surface effects and all interacting molecules are explicity included.
However, for single monolayers or bilayers the periodic boundary conditions
should be applied in two directions, in the plane of the mono- or
bilayers ( \emph{x y} plane), but not along the perpendicular to the
plane (\emph{z} axis). 2D Ewald sums have been developed for these
cases, although they are computationally very lenghly by comparison
with the 3D sums \cite{jorge-2d,ewald-2d-deleeuw}. Recently Brôdka
and Grzybowski \cite{ewald-mz2} showed analytically in which way
the 2D Ewald sums may be accurately approximated by a 3D calculation.
They show that, in order to obtain a reliable calculation, not only
a large empty space must be introduced in the simulation box (along
\emph{z}), but the macroscopic electric field term depending on the
total dipole moment $\mathbf{M}$ of the sample ($\left|\mathbf{M}\right|^{2}/3)$
should be replaced by a term containing only the component $z$ perpendicular
to the 2D system ($M_{z}^{2}$). In ref. \cite{ewald-mz} this result
was checked with a numerical simulation. It has to be taken into account
that, in a MD simulation, the contribution of this macroscopic electric
field to the molecular interactions fluctuates in time and is far
from negligible \cite{ewald-mz}, particularly for monolayers.\\

The core of the paper is in Section IV where we propose a novel, simple
and more accurate macroscopic field for model bilayers. That is, a
model that goes beyond the total dipole moment of the sample, which
on time average is zero for this type of symmetrical bilayers. Following
this approach, we extend it by including higher order moments of this
macroscopic electric field. We show that by representing it with a
superposition of gaussians it can be \emph{analytically} integrated,
and therefore its calculation easily implemented in a MD simulation
(even in simulations of non-symmetrical bilayers). \\

In this paper we study a simple model of Newton black films via MD
simulations and using the TIP5P intermolecular potential for water
and a flexible charged chain to simulate the amphiphilic molecules.
We show that our method of calcutation of the macroscopic electrostatic
potential reproduces the spatial and temporal charge inhomogeneities
of the quasibidimensional sample and is extremely simple to implement
in numerical simulations. We also show that this method can be directly
applied to any other type of bi- or multi-layers.

\section{The amphiphilic and water molecule models}

Our simple model of a charged amphiphilic consists in a semiflexible
single chain of 14 atoms, the bond lengths are considered constant,
but bending and torsional potentials are included. The first two atoms
of the chain mimic a charged and polar head (atom 1 with $q_{1}$
=- 2 e, atom 2 with $q_{2}$ = 1 e), sites 3 to 14 form the hydrophobic
tail with uncharged atoms: sites 3 to 13 are uncharged united atom
sites $CH_{2}$ and site 14 is the uncharged united atom site $CH_{3}$.
This molecular model correspond to an oversimplification of sodium
dodecyl sulfate SDS ($CH_{3}(CH_{2})_{11}OSO_{3}^{-}\, Na^{+}$) in
solution, so we are including, in our simulations,  a $Na^{+}$ ion
per chain. The LJ parameters of these interaction sites are those
of ref. \cite{zg}, except for the sites 1 and 2 that form the amphiphilic
polar head: $\sigma_{1}=4.0$$\textrm{Å}$, $\sigma_{2}=$4.0 $\textrm{Å}$,
$\sigma_{Na}=$1.897 $\textrm{Å}$, $\varepsilon_{1}$= 2.20 kJ/mol,
$\varepsilon_{2}$= 1.80 kJ/mol and $\varepsilon_{Na}$= 6.721 kJ/mol.
The masses of the sites are the corresponding atomic masses, except
that $m_{1}=m_{2}=48$au., in order to mimic the 'real' amphiphilic
head. The LJ parameters of the united atom sites are taken from calcutations
on \emph{n-}alkanes \cite{pot.toxvaerd}: $\sigma_{CH2}=$3.850 $\textrm{Å}$,
$\sigma_{CH3}=$3.850 $\textrm{Å}$, $\varepsilon_{CH2}$= 0.664 kJ/mol,
$\varepsilon_{CH3}$= 0.997 kJ/mol. The LJ parameters for the $Na^{+}$
ion are taken from simulations of SDS in aqueous solution \cite{mike-sds-miscelle}
and NB films \cite{zg}. \\

The intramolecular potential includes harmonic wells for the bending
angles $\beta$ and the usual triple well for the torsional angles
$\tau$, the constants are those commonly used for the united atom
site $CH_{2}$\cite{tildesley}. These potentials are needed to maintain
the amphiphilic stiffness and avoid molecular collapse. The bending
potential is\[
V(\beta)=k_{CCC}(\beta-\beta_{0})^{2}\;,\]
with $\beta_{0}=109.5\, deg.$ and $k_{CCC}=520\, kJ/rad^{2}$. The
torsional potential is of the form

\[
V(\tau)=a_{0}+a_{1}cos(\tau)+a_{2}cos^{2}(\tau)+a_{3}cos^{3}(\tau)+a_{4}cos^{4}(\tau)+a_{5}cos^{5}(\tau)\;,\]
the constants are $a_{0}=9.2789,$ $a_{1}=12.1557,$ $a_{2}=-13.1202,$
$a_{3}=-3.0597,$ $a_{4}=26.2403$ and $a_{5}=-31.4950\, kJ/mol$;
this potential has a main minimum at $\tau=0\, deg.$ and two secondary
minima at $\tau=\pm120\, deg.$\\

The selected molecular model for water is the classical rigid molecular
model TIP5P \cite{jorgensen1,jorgensen2}. It consists of one LJ site
($\varepsilon=0.67$ kJ/mol, $\sigma=$ 3.12 $\textrm{Å}$) localized
at the O and 4 charges. Two charges q$_{\textrm{H}}=0.241e$ are localized
at the H atoms and two q$_{Lp}=$-q$_{\textrm{H}}$  at the lone pairs.
As the lone pair interaction sites are not coincident with atomic
sites, the algorithm employed to translate the forces from massless
to massive sites is that of ref. \cite{algor6}. The final version
of the MD program is similar to that used in refs. \cite{cyclobutane,sulfur-jcp01,sulfur-jcp2003}.

This rigid, nonpolarizable, TIP5P molecular model was selected by
us because it is simple and it also gives good results for the calculated
energies, diffusion coeficient and density $\rho$ as a function of
temperature, including the anomaly of the density near 4C and 1atm
\cite{jorgensen1}, the X-ray scattering of liquid water \cite{parrinello-w},
etc. The only exception is the O-O pair correlation function $g_{2}(r)$,
for which the first neighbor  is located at a slightly shorter distance
than the experimental one \cite{jorgensen1}. Very recently a six
sites rigid model for water \cite{tip6p}, with an additional site
at the molecular center of mass, has been presented. By comparison
with the TIP5P model, the new model has improved the fit of the melting
point and the disordered structure of ice at the melting point. Nevertheless,
this six sites model still shows the same problem of TIP5P to reproduce
the experimental $g_{\textrm{OO}}(r)$ pair correlation function and
this is the reason to perform our calculations with TIP5P.

\section{The model bilayer of amphiphilic molecules:}

\begin{figure}
[!ht]\includegraphics[width=0.25\paperwidth,angle=90]{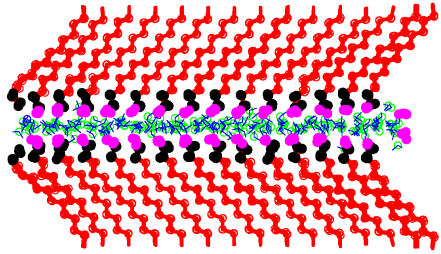}

\caption{Initial configurations of a model Newton black film.\label{cap:mem226c14-iniconf}}
\end{figure}

Fig. \ref{cap:mem226c14-iniconf} shows the initial configuration
of one of our simple model of a NB film (i. e. soap bubbles films)
with the water within the bilayer and, therefore, the head groups
oriented to the inside. The sample of the figure includes 226 amphiphilics
and 365 water molecules, and, as in model (a), the bilayer is perpendicular
to the $\widehat{z}$ MD box axis, with a box size of $a=b=45.$$\textrm{Å}$,
$c=$1000 $\textrm{Å}$. 

The periodic boundary conditions are applied only in the $\widehat{x}$
and $\widehat{y}$ direction and a large unit cell $c$ parameter
(that is, a large empty volume) is taken, in order to approximate
the required 2D Ewald sums by the usual 3D sums plus our macroscopic
field term corresponding to a quasi-2D system, discused in the following
section. \\

When performing these simulations, we have to take into account, in
particular, the following facts:

a) The site-site LJ interactions between all molecules in the sample
have a finite cutt-off radius of 15 $\textrm{Å}$. In 3D simulatons,
the contributions of sites outside this sphere are taken into account
assuming an uniform distribution of sites and performing a simple
integration. In our cuasi 2D system, the volume to integrate is that
outside the cut-off sphere and within the volume of the slab. Appendix
A includes this integral.

b) The electrostatic interactions are calculated \emph{via} the standard
3D Ewald sums with a large size box along the perpendicular to the
bilayer slab and 2D periodic boundary conditions in the plane of the
slab. The Ewald's sum term corresponding to the macroscopic electric
field is discussed in the following section.

\section{The macroscopic electric field in a quasi - 2 D sample:}

Electrostatic forces have a far from negligible contribution to the
self-assembly and final patterns found in soft matter systems. For
bilayers the macroscopic electric field is given, in a first approximation,
by the contribution of the surface charges of the macroscopic dielectric
slab. There are very clear reviews where the first multipolar moment
of these charges are taken into account, in 2D and 3D macroscopic
samples \cite{ewald-2d-deleeuw,ewald-mz}. In particular, for an uniform
dielectric slab oriented perpendicular to the z direction, the contribution
of the uniformly distributed surface charges to the total energy of
the macroscopic slab system, of volume  $V$, is:

$U^{macrosc.}=\frac{2\pi}{V}M_{z}²$, \\
where $M_{z}$ is the total dipole moment of the slab. Then the contribution
of this term to the total force on every charge $q_{i}$ of the sample
is:

$F_{i}^{macrosc.}(z)=-\frac{4\pi}{V}M_{z}$.\\
This approximation has been tested in monolayers \cite{ewald-mz2,ewald-mz}.
In the MD simulation of bilayers, and due to its geometry, the total
dipole moment $M_{z}$ is zero in a time average. Nevertheless, as
in monolayers the forces $F_{i}^{macrosc.}(z)$ are not negligible
by comparison with the rest of the interaction forces, and, as the
electrostatic interactions play a key rôle in the dynamics and structural
properties of the bilayers, a more accurate estimation of their macroscopic
electric field is desirable. One possibility is to include higher
multipolar moments of the surface charges, but the convergence of
the electrostatic interactions using total multipole series is slow.
Another possibility is to solve the Poisson-Boltzman equation with
boundary conditions given by the surface charges \cite{andelman},
this method is extremely lenghtly for a numerical simulation, because
the charged atoms are mobile and change their location in every time
step.\\

Nagle \& Nagle \cite{naglenagle,nagle2} recently reviewed the experimental
data on the structure of lipid bilayers, in particular the distribution
functions for the symmetrical components along the direction of the
normal to the bilayer. They analize not only the peak positions of
each distribution but their shape, pointing out that they are slightly
asymmetrical (to the interior or exterior of the bilayer) and therefore
a gaussian distribution model is just an useful first approximation
when only the positions and widths are available. \\

Here we propose a novel coarsed fit for the charge distribution of
the different bilayer components (water and charged amphiphilics plus
ions) using a superposition of gaussian distributions. We found that,
in this way, the contribution of these charge distributions to the
macroscopic electric field can be exactly calculated. The method is
extremely simple to implement in numerical simulations, and the spatial
and temporal charge inhomogeneities are roughly taken into account.
Our results can also be applied to biological membrane models.

At each time step of the MD simulation we decompose our bilayer's
charge distribution in four neutral slabs: two for the upper and lower
water layers and other two for the neutral layers formed by the head
plus ion charges. 

For each one of the four neutral slabs, instead of consider two planar
surfaces with an uniform density of opposite charges, we propose two
gaussian distribution along \emph{z}, the perpendicular to the slabs,
with the same opposite total charges and located the same relative
distance (maintaining the slab width). The coarsed distribution of
charges in the bilayer is then a linear superposition of gaussians:\\

$\rho(z)=\sum_{i}\,\frac{q_{i}}{\sqrt{2\pi}\sigma_{i}}\exp(-\frac{(z-z_{i})²}{2\sigma²_{i}}))\quad,$\\

The macroscopic electric potential $V(z)$ and the force field $E_{z}(z)$
due to this type of charge distribution can be exactly solved. In
Appendix B we include the analytically solved integrals (one of them
is a new mathematical solution). The final result is:\\

$V(z)=-2\sqrt{2\pi}\sum_{i}\,\sigma_{i}\, q_{i}\,(\exp(-\frac{(z-z_{i})²}{2\sigma_{i}²})-(\frac{z-z_{i}}{\sqrt{2}\sigma_{i}})\, Erf[\frac{(z-z_{i})}{\sqrt{2}\sigma_{i}}])\,.$

$E_{z}(z)=-\frac{\partial V(z)}{\partial z}=2\pi\sum_{i}\, q_{i}\, Erf[\frac{(z-z_{i})}{\sqrt{2}\sigma_{i}}].$\\

The obtained result is valid for any number of slabs, that as a function
of time can change not only their position and width but also they
can superpose. The practical implementation of this macroscopic field
in our simulations (that is, how the $\sigma_{i},$ $z_{i}$ and $q_{i}$
values are taken at each time step) is explained in the following
section. \\
The contributions of this macroscopic field, as well as that of the
external field $U_{eff}(h)$ in biological membranes, calculated for
several bilayers samples are included in section 7. Units: density
of charges $\left[\rho\right]=\frac{e}{\textrm{Å}^{3}}$; electrostatic
potential $\left[V\right]=\frac{e}{\textrm{Å}}$ (for comparisson
with experimental data $\left[\frac{e}{\textrm{Å}}\right]=0.04803\,\frac{cm^{1/2}gr^{1/2}}{sec}=14.399\,\left[Volt\right]$
); electric field $\left[E\right]=\frac{e}{\textrm{\textrm{Å}}^{2}}$.\\

Here we have applied this exact calculation method of the macroscopic
electric field to a symmetrical (along \emph{z}) slab geometry, but
it is also valid in an asymmetrical case, which may imply a finite
difference of potential across the bilayer. 

The extension of this method ( a coarse grained representation of
the macroscopic electric field \emph{via} a superposition of gaussians)
to other geometries is also strightforward, a spherical geometry,
for example, would be useful for the study of miscelles. Its great
advantage is that these representations can be a\emph{nalytically}
solved. Moreover, although the present work is not dedicated to the
study of reaction fields, based on a coarse gaussian distribution
of charges, can be also be easily applied to include a more realistic
reaction field inside a cavity in non-uniform dielectrics.

\section{Implimentation of the numerical simulations:}

In our MD simulations of the bilayers of Fig. \ref{cap:mem226c14-iniconf},
the integration algorithms, time step and cut-off radius are essentially
identical to those used on our bulk samples \cite{cyclobutane,sulfur-jcp2003},
except for the periodic boundary conditions, that now are applied
only in the xy plane of the bilayers, and that the calculations include
now our proposed macroscopic electric field. The equations of motion
of the rigid water molecules are integrated using the velocity Verlet
algorithm for the atomic displacements and the Shake and Rattle algorithms
for the constant bond length constraints on each molecule. The time
step is of 1 fs., the sample is thermalized for 20 ps. and measured
in the followings 100 ps. \\

To maintain a constant temperature in our simulations of these bilayers,
the Nosé-Hoover chains method \cite{nose-chains} (that we previously
used for bulk samples) was disregarded because of the lenghly termalization
of all the components, in many cases we obtained a different equilibrium
temperature for each different molecular species. Instead, we chose
to perform our MD simulations of the bilayers using the Berendsen
algorithm \cite{berendsen1}, applying the equipartition theorem to
each type of molecule and a strong coupling constant $\tau=0.5\,\Delta t$
linking the average kinetic energy of each kind of molecules (amphiphilics,
ions and water) to the desired kinetic energy of $\frac{3}{2}k_{B}T_{0}$,
with $T_{0}=300K$. The Berendsen algorithm attains equipartition
and equilibrium temperatures faster than the Nose-Hoover chains method,
when applied to our mix of flexible and rigid molecules.\\

To include the macroscopic electric field term we need to determine
the values of the $\sigma_{i},$ $z_{i}$ and $q_{i}$ parameters
at each MD time step. For each one of the two slabs that simulate
the charge distribution of chains' heads plus ions, we fit the $z_{i}$
parameters of two gaussians, so as to reproduce the dipolar moment
of the slab. Their $\sigma_{i}$ values are obtained from the corresponding
charge distributions, with $q_{i}=1$ for ions and $q_{i}=-1$ for
chains' heads. Typical values of these variables, as well as the contribution
of the external potential and the macroscopic electric field to the
total forces on all molecules, are reported for each calculated case.

\section{results:}

Here we present the results obtained for one sample of our simple
NB film model. Our simulations were mainly performed to analyze the
contribution of the macroscopic electric fields to the equilibrium
structure and molecular dynamics of the bilayer. Elsewhere we will
test the versatility of our approach by studying biological membrane
models with and without ions solved in water, with and without the
water layer, as a function of chain length, etc..

\begin{figure}
[!ht]\includegraphics[width=0.25\paperwidth,angle=90]{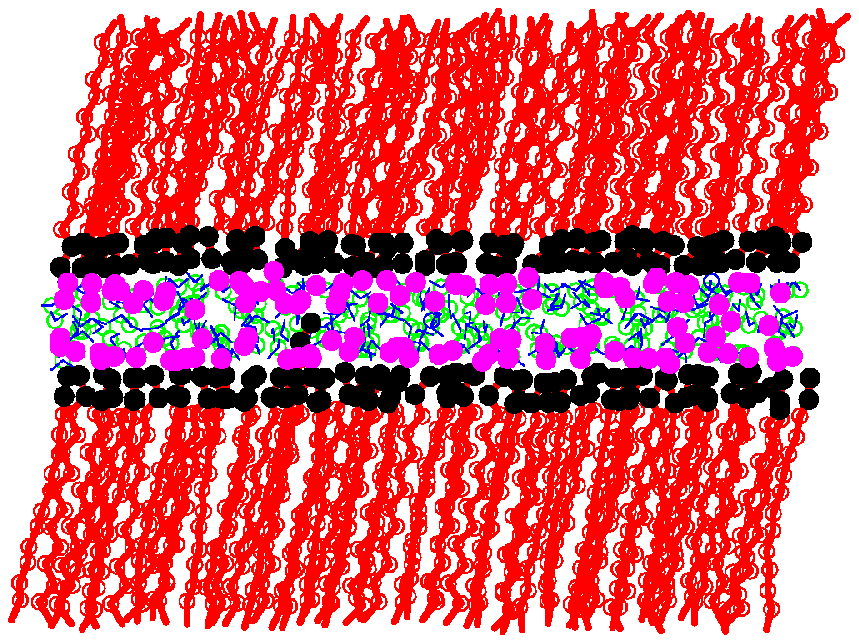}$\quad$\includegraphics[width=0.27\paperwidth]{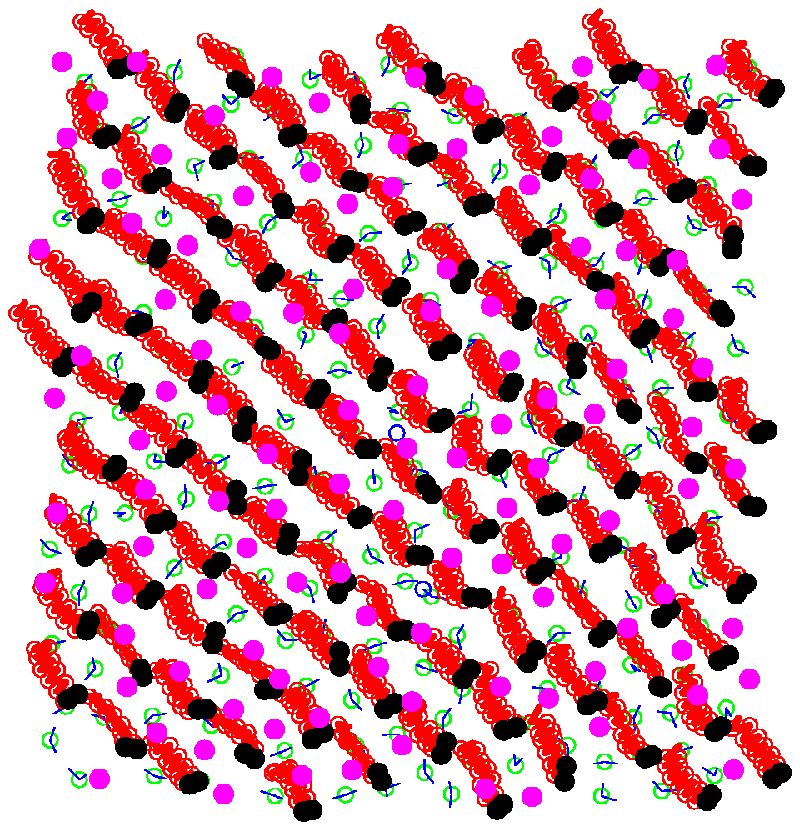}

\caption{Final configuration of the HD sample with a=b=45$\textrm{Å}$: (a)
\emph{ac
} cross section, (b) \emph{ab} cross section.\label{cap:Final-conf.soap.bub.a45}}
\end{figure}

In Ref. \cite{zg} a MD simulation with 64 all-atom sodium-dodecyl-sulfate
SDS {[}Na$^{\textrm{+}}$CH$_{\textrm{3}}$(CH$_{\textrm{2}}$)$_{\textrm{11}}$(OSO$_{\textrm{3}}$)$^{\textrm{-}}$]
molecules and 1 to 6 water molecules per amphiphilic was presented.
Here we use our simplified amphiphilic model. In section 4, Fig. \ref{cap:mem226c14-iniconf}
\emph{(b)} showed the initial configuration of a sample consisting
of 226 charged amphiphilics and 226 $Na^{+}$ ions solved in 365 water
molecules, the bilayer is perpendicular to the $\widehat{z}$ MD box
axis, with a box size of $a=b=45$.5$\textrm{Å}$, $c=$1000 $\textrm{Å}$.
The film is thermalized for 50 ps. and measured in the following 100
ps., Fig. \ref{cap:Final-conf.soap.bub.a45} shows the attained final
configuration of this sample (HD high density), two other samples
with a lower density of amphiphilic heads per unit area were calculated:
a medium density MD ($a=b=46.5$$\textrm{Å}$) and a LD low density
sample ($a=b=48.0$$\textrm{Å}$). \\

\begin{figure}
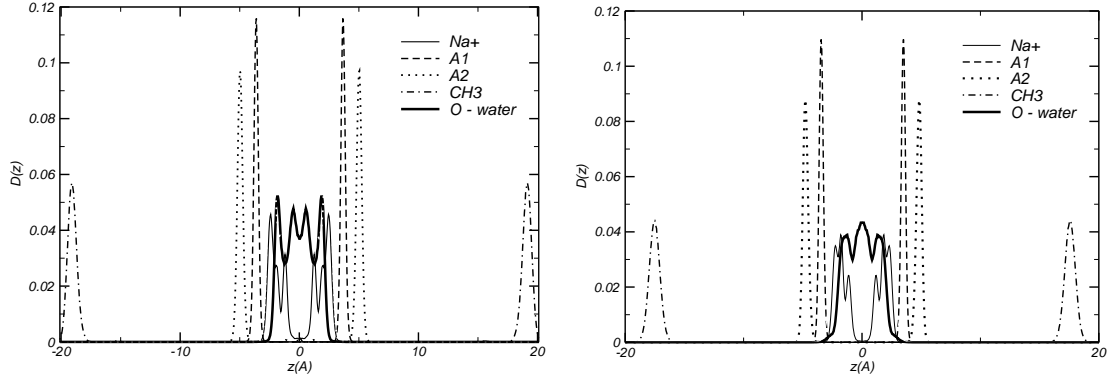

\includegraphics[width=0.34\paperwidth]{soapbub-wiamp14-365w-aug3006-a50-alg26-a9-zg2-otra-densa}$\quad$\includegraphics[scale=0.3]{soapbub-wiamp14-365w-aug3006-a48-alg25-a17-densa}

\caption{Atomic density profiles (number of atoms per $\textrm{Å}³$) for
the (\emph{a}) HD and (\emph{b}) LD head group samples.\label{cap:soap.bub.Atomic-dens-profiles}\protect \\
}
\end{figure}

The width of our HD film is 38.2$\textrm{Å}$, of our MD film is 36.8$\textrm{Å}$
and that of the LD one is 35.1$\textrm{Å}$, as measured from the
average distance between end tail $CH_{3}$ groups. Fig. \ref{cap:soap.bub.Atomic-dens-profiles}
shows the atomic density profiles in two cases. X-ray reflectivity
measurements determined that common soap films (CBF) are several thousand
angstroms wide, while a NBF is about 33 $\textrm{Å}$ \cite{soap.bub.thick0}.
In ref. \cite{soap.bub.thick} a dual optical multiwave interferometer
( of $\lambda=1.064\mu m$) was used to determine the dynamics of
gravity induced gradients in soap film thicknesses, which allows to
detect variations of about 1$\textrm{Å}$ in the thickness of the
film.

\begin{figure}
\includegraphics[scale=0.4]{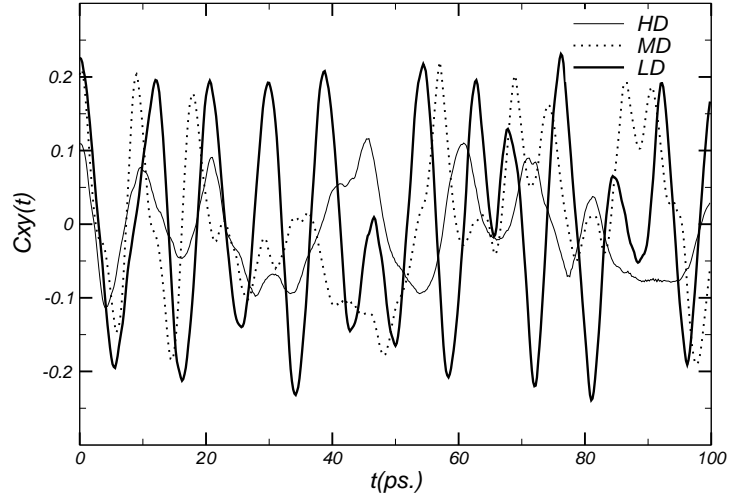}

\caption{Reorientational autocorrelation in the \emph{xy} plane.\label{cap:soap.reorxy}\protect \\
}
\end{figure}

The three cases correspond to a glassy phase, as measured by their
diffusion coefficients. For the HD sample the molecular centers of
mass location show average oscillations with an amplitude of $\sim$3
$\textrm{Å}$ and we measured an overall displacement of about 5 $\textrm{Å}$
in 100 ps., within the bilayer plane, therefore our measured diffusion
coefficient is less than our measurement error (about $10^{-6}cm^{2}/sec$).
The experimental value of the diffusion coefficient for SDS, in a
film of about 35$\textrm{Å}$ thickness, is $6\:10^{-7}cm^{2}/sec$
\cite{soap.bub.diff,soap.bub.proteins2} . In the MD and LD samples
we observe a collective rotation or large amplitud oscillation of
the amphiphilic molecules around their corresponding center of mass
and about an axis perpendicular to the bilayer plane. We measure a
tilt angle of 26.2 deg. in our HD sample, 35.9 deg. in the MD and
40.2 deg. in the LD sample. Fig. \ref{cap:soap.reorxy} shows the
autocorrelation function of the $xy$ component of the reorientational
molecular motion, showing a disordered reorientation at HD and an
oscillatory motion at LD.\\

\begin{figure}
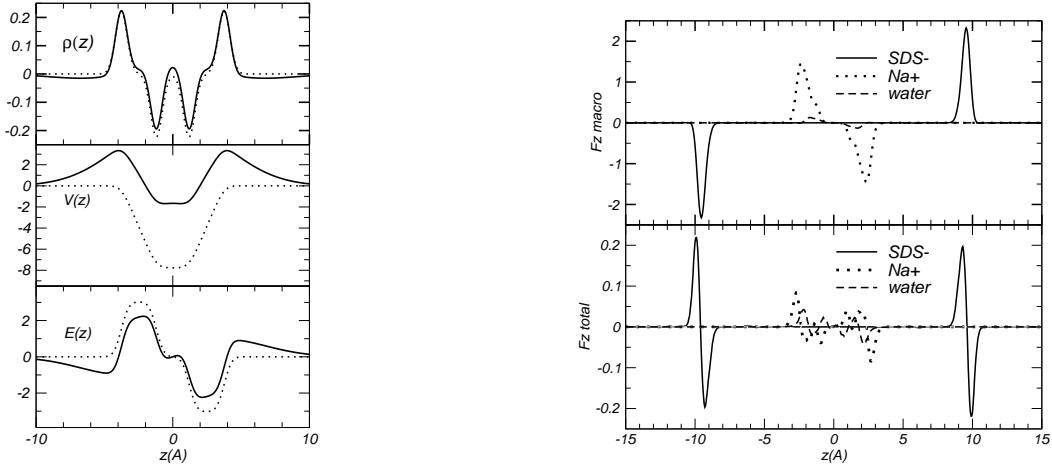

\includegraphics[width=0.2\paperwidth]{soapbub-wiamp14-365w-aug3006-a50-alg26-a9-vself-tot-w}$\quad\quad\quad\quad\quad\quad\quad\quad\quad$\includegraphics[width=0.3\paperwidth]{soapbub-wiamp14-365w-aug3006-a50-alg26-a9-zg2-fz-mido}

\caption{Profile of: (a) the macroscopic charge density $\rho$(z), electric
field and potential (full line for total values, dotted line for water
contribution), (b) our measured profile of macroscopic electric and
total forces on molecules, as a function of the location of their
centers of mass along $z$, in the HD, NB film. \label{cap:soap.bub.Profile.fz.medido}}
\end{figure}

As in the previous bilayers models, the funcion $\rho(z)$ is obtained
from a coarse grained fit of the charge distribution of our MD sample
using four charged slabs. The parameters that describe these slabs
are for the LD film are:

$\begin{array}{lrrr}
 & q_{i}(e) & z_{i}(\textrm{Å}) & \sigma_{i}(\textrm{Å})\\
slab\,1\,(head\, and\, ions) & -1.0 & 3.485 & 3.344\\
 & 1.0 & 2.733 & 3.344\\
slab\,2\,(water) & -0.241 & 1.196 & 0.433\\
 & 0.241 & 3.762 & 0.433\\
slab\,3\,(water) & 0.241 & -3.762 & 0.433\\
 & -0.241 & -1.196 & 0.433\\
slab\,4\,(head\, and\, ions) & 1.0 & -2.733 & 3.344\\
 & -1.0 & -3.485 & 3.344\end{array}$$\quad,$\\
and with them we calculate the functions included in Fig. \ref{cap:soap.bub.Profile.fz.medido}
\emph{(a)} . As this is a very thin NB film with about 1.6 water molecules
per amphiphilic, water is highly packed and strongly polarized due
to the electric field generated by the charged head and ions. In Fig.
\ref{cap:soap.bub.Profile.fz.medido} \emph{(a)} we also included
the strong contribution of the water layer to the calculated profiles
of our samples. This behavior has a strong dependence on the content
of water in the NBF and will be presented elsewhere. Our calculated
total electrostatic potential is nearly constant at the core, with
a negative value of about -1.69 $e/\textrm{Å}$. Although not directly
comparable, in Ref. \cite{soap.bub.dielec} thicker NB films with
up to 11.94 water molecules per amphiphilic were studied with an all-atom
MD simulation and also a strong polarization of water was found at
the interface with the head layers. Their measured electrostatic potential
is also nearly constant at the core, but with a much smaller negative
value of about -0.2 Volt= -0.0139$e/\textrm{Å}$ for the sample of
11.94 water molecules per amphiphilic. Unfortunately there is not
experimental data available for comparison, but nevertheless, in Ref.
\cite{soap-faraudo-05} the origin of the short-range and strong repulsive
force between two ionic surfactant layers is calculated as decaying
exponentially with their distance.\\

Fig. \ref{cap:soap.bub.Profile.fz.medido} \emph{(b)} includes the
contribution of macroscopic electric field $E$ to our measured total
forces on water, ions and chain molecules, as a function of the centers
of mass location, averaged over a MD free trajectory of 100 ps. Units:
$\left[F\right]=kJ/mol/\textrm{Å}$. Due to the macroscopic electric
field, the negatively charged amphiphilics tend to drift away of the
bilayer but this tendency is balanced by a contrary force on the movile
ions, that remain within the water layer. When adding all molecule
- molecule interactions, the total forces (an order of magnitude lower
than those due to the macroscopic field) on the amphiphilics tend
to maintain a bilayer structure with the water inside. \\

\section{Conclusion:}

In this paper we studied a simple model of a amphiphilic bilayer,
a Newton black film, and analyzed the key \emph{rôle} of the electrostatic
interactions in their self-assembly. The model is simple enough so
a large number of molecules can be included in the MD samples, but
also detailed enough so as to take into account molecular charge distributions,
flexible amphiphilic molecules and a reliable model of water. All
these properties are essential to obtain a reliable conclusion.\\

The presented amphiphilic molecular model is extremely simple, but
includes all main characteristics of these type of molecules: a strongly
charged head atoms and a tail consisting in a semiflexible chain of
hydrocarbon united atoms. As an example of its versatility we studied
samples with a tail of 12 atoms and other of 18 atoms. This amphiphilic
model also allows to study, in a simple way, the properties of bilayers
formed by charged or neutral amphiphilics and with or without explicity
including water molecules in the numerical simulations.\\

As for the calculation of the electrostatic interactions, we also
propose a novel and more accurate method to calculate the macroscopic
electric field in cuasi 2D geometries, which can be easily included
in any numerical calculation. The method, that essentially is a coarse
grain fit of the macroscopic electric field beyond the dipole order
approximation, was applied here to symmetrical bilayers (along the
normal to the bilayer and with periodic boundary conditions in two
dimensions), but their derivation is general and valid also for asymmetrical
slab geometries.\\

Lastly, we emphasize the relevance and utility of these simple bilayer
models. They can be applied to the systematic study of the physical
properties of these bilayers, that strongly depend not only on 'external
parameters', like surface tension and temperature, but also on the
kind of guest molecules embedded in them (eg. proteins, etc.). In
turn, the structure and dynamics of the embedded molecules strongly
depend on their interactions with the bilayer and with the sorrounding
water. 

These simple bilayers can also be extremely useful to model, for example,
the synthesis of inorganic (ordered or disordered) materials \emph{via}
an organic agent, which are based on the use of these self-assembling
molecules \cite{nanotech}. This is a recent and very fast growing
research field of nanotechnologycal relevance, as are lithography,
etching and molding devices at the nanoscopic scale. Another fast
developing field is that of electronic sensors and nanodevices supported
on lipid bilayers \cite{nanotech2,nanotech3}.

Furthermore, the amphiphilic bilayers are hosts for the diffusion
and/or nucleation of guest molecules of relevance in biological and/or
ambient problems, these studies are currently being performed. 

Lastly, as our model retains the flexibility of the original amphiphilics,
and the electrostatic interactions are included, the approach is really
useful to obtain 'realistic' solutions to the above mentioned problems
as well as those related to electric fields and electrostatic properties.\\

\begin{acknowledgments}
Z. G. greatly thanks careful reading and helpful sugestions to J.
Hernando.
\end{acknowledgments}

\section*{\textbf{A: Correction terms for the atom-atom interactions within
the slab and beyond the cut-off radius} }

In 3D samples, correction terms to the energy $U^{3D}(R_{cut})$ and
virial $Vir^{3D}(R_{cut})$, due to the finite cut-off radius of the
LJ site-site interactions, are calculated asuming an uniform distribution
of sites beyond the cut-off radius $R_{cut}$. These terms are:\\

$U^{3D}(R_{cut})=\int_{0}^{\pi}\sin\theta\, d\theta\int_{R_{cut}}^{\infty}U_{LJ}(r)\,2\pi r^{2}\, dr=\frac{4\pi}{3}\varepsilon\sigma^{3}\left[\frac{1}{3}\left(\frac{\sigma}{R_{cut}}\right)^{9}-\left(\frac{\sigma}{R_{cut}}\right)^{3}\right],$
and

$Vir^{3D}(R_{cut})=\int_{0}^{\pi}\sin\theta\, d\theta\int_{R_{cut}}^{\infty}(-r\frac{\partial U_{LJ}(r)}{\partial r})\,2\pi r^{2}\, dr=8\pi\varepsilon\sigma^{3}\left[\frac{2}{3}\left(\frac{\sigma}{R_{cut}}\right)^{9}-\left(\frac{\sigma}{R_{cut}}\right)^{3}\right].$\\

In our 2D samples, instead, the integrals are over the volume outside
the $R_{cut}$ radius and inside the slab. In all of our cases the
width of the slab $z_{slab}$ is larger than the diameter of the cut-off
sphere $(z_{slab}>2R_{cut}),$ and the correction terms are then simply
calculated. 

$U^{2D}(R_{cut})=\int_{0}^{\pi}\sin\theta\, d\theta\int_{R_{cut}}^{R_{max}(\theta)}U_{LJ}(r)\,2\pi r^{2}\, dr$,
\\
where $\theta$ is the azimutal angle measured from the perpendicular
to the bilayer and $R_{max}(\theta)=\frac{z_{slab}}{2\cos(\theta)}$.

The final result is: 

$U^{2D}(R_{cut})=U^{3D}(R_{cut})-\frac{4\pi}{3}\varepsilon\sigma^{3}\left[\frac{1}{30}\left(\frac{2\sigma}{z_{slab}}\right)^{9}-\frac{1}{4}\left(\frac{2\sigma}{z_{slab}}\right)^{3}\right]$.

In a similar way we obtain:

$\begin{array}{ll}
Vir^{2D}(R_{cut}) & =\int_{0}^{\pi}\sin\theta\, d\theta\int_{R_{cut}}^{R_{max}(\theta)}(-r\frac{\partial U_{LJ}(r)}{\partial r})\,2\pi r^{2}\, dr\\
 & =Vir^{3D}(R_{cut})-4\pi\varepsilon\sigma^{3}\left[\frac{4}{30}\left(\frac{2\sigma}{z_{slab}}\right)^{9}-\frac{1}{2}\left(\frac{2\sigma}{z_{slab}}\right)^{3}\right],\end{array}$\\
and for the component along $z$ (used to control the pressure along
the perpendicular to the bilayer):

$\begin{array}{ll}
Vir_{z}^{2D}(R_{cut}) & =\int_{0}^{\pi}\sin\theta\, d\theta\int_{R_{cut}}^{R_{max}(\theta)}(-z\frac{\partial U_{LJ}(r)}{\partial z})\,2\pi r^{2}\, dr\\
 & =\frac{1}{3}Vir^{3D}(R_{cut})-\frac{4\pi}{3}\varepsilon\sigma^{3}\left[\frac{1}{3}\left(\frac{2\sigma}{z_{slab}}\right)^{9}-\left(\frac{2\sigma}{z_{slab}}\right)^{3}\right].\end{array}$\\

On the other hand, if $z_{slab}<2R_{cut}$ , the calculated terms
are:

$U^{2D}(R_{cut})=U^{3D}(R_{cut})\,\frac{z_{slab}}{2R_{cut}}-\frac{4\pi}{3}\varepsilon\sigma^{2}z_{slab}\left[\frac{1}{30}\left(\frac{\sigma}{R_{cut}}\right)^{9}-\frac{1}{4}\left(\frac{\sigma}{R_{cut}}\right)^{3}\right]$, 

$Vir^{2D}(R_{cut})=Vir^{3D}(R_{cut})\,\frac{z_{slab}}{2R_{cut}}-4\pi\varepsilon\frac{z_{slab}}{R_{cut}}\sigma^{3}\left[\frac{4}{30}\left(\frac{2\sigma}{z_{slab}}\right)^{9}-\frac{1}{2}\left(\frac{2\sigma}{z_{slab}}\right)^{3}\right]$,
and 

$Vir_{z}^{2D}(R_{cut})=(\frac{z_{slab}}{2R_{cut}})^{3}\left[\frac{1}{3}Vir^{3D}(R_{cut})-U^{3D}(R_{cut})\right].$

\section*{\textbf{B: The electrostatic macroscopic field model} }

We propose a superposition of slabs, upper and lower surfaces with
opposite charges, but with a gaussian distribution, of width $\sigma$,
along the depth of the slab. The charge distribution, electrostatic
potential and electric field expresions are given in the following
subsections, in order of increasing complexity. Case C.2 is the one
we use for the reaction field in our MD simulations.

The results showed here correspond to a symmetrical distribution of
charges, in the bilayer, about the origin (i. e. identical charges,
widths and location of the maximums along $\pm$\emph{z} axis. But
for an asymmetrical case the calculation is straightforward, and the
obtained result should be very useful for studying these bilayers
under an applied external electric field, for example.

\subsection*{B1: The macroscopic electric field of a single slab: }

First of all we review the well known case of an uniformly charged
surface ($\rho$ per unit area) located at $z=0$. The potential field
at $z$, due to this infinite, uniformly charged (\emph{xy}) plane
is:

$V(z)=\int_{0}^{\infty}\rho\frac{2\pi rdr}{\sqrt{z²+r²}}=2\pi\rho\left[\sqrt{x}\right]_{z²}^{\infty}$,

and the electric field is:

$E(z)=-\frac{\partial V(z)}{\partial z}$= 2$\pi\rho\frac{z}{\left|z\right|}.$\\

For a homogeneous dielectric single slab (of width $2z_{i}$), the
macroscopic electric field is calculated assuming an uniformly charged
plane at $z_{i}$ ($q_{i}$ per unit area) and another with opposite
charge ($-q_{i}$ per unit area) located at -$z_{i}$. The potential
field of this charge distribution is:

$V(z)=2\pi q_{i}\,(\left[\sqrt{x}\right]_{(z-z_{i})²}^{\infty}-\left[\sqrt{x}\right]_{(z+z_{i})²}^{\infty})=-2\pi q_{i}(\left|z-z_{i}\right|-\left|z+z_{i}\right|)$,
that is,\\

$V(z)=\left\{ \begin{array}{rcr}
-4\pi q_{i}z_{i} & if & z<-z_{i}\\
4\pi q_{i}z\, & if & -z_{i}\leqslant z\leqslant z_{i}\\
4\pi q_{i}z_{i} & if & z_{i}>z\end{array}\right.$.

The electric field of this charge distribution is:

$E(z)$=-4$\pi q_{i}$, if$-z_{i}\leq z\leq z_{i},$ and $E(z)=0$
if $z<-z_{i}$ or $z>z_{i}.$

Usually this result is given in terms of the the total dipole moment
of the slab per unit area, which is $M_{z}=2q_{i}z_{i}$, the macroscopic
electric field then is $E_{z}=-\frac{2\pi}{z_{i}}M_{z}$, and the
contribution of this charge distribution to the total energy of the
system is $\frac{\pi}{z_{i}}M_{z}^{2}$.\\

If instead of considering two uniformly opposite charged planes, we
consider that the charges show a certain spread along \emph{z}, we
can analyze a model of two gaussian charge distributions along \emph{z},
of total charge $\pm q_{i}$ (per unit area) each one, standard deviation
$\sigma_{i}$ and maximum located at $\pm z_{i}$respectively, and
the charge distribution across the slab becomes now:

$\rho(z)=\frac{q_{i}}{\sqrt{2\pi}\sigma_{i}}(\exp(-\frac{(z-z_{i})²}{2\sigma²_{i}})-\exp(-\frac{(z+z_{i})²}{2\sigma²_{i}}))\quad.$

The total dipole moment of this system is again $M_{z}=2q_{i}z_{i}$,
but the potential $V(z)$ will differ. For the sake of simplicity
we take as equal the dispersion of both distribution and we obtain:\\

$V(z)=-2\pi\int_{-\infty}^{\infty}\rho(t)\left|z-t\right|dt=-2\pi(\int_{-\infty}^{z}\rho(t)(z-t)dt+\int_{z}^{\infty}\rho(t)(t-z)dt)$.\\
Replacing the $\rho(t)$function we first obtain:\\
 $\begin{array}{ll}
V(z)= & -\sqrt{2\pi}\frac{q_{i}}{\sigma}\,(\int_{-\infty}^{z}\exp(-\frac{(t-z_{i})²}{2\sigma²})(z-t)dt-\int_{-\infty}^{z}\exp(-\frac{(t+z_{i})²}{2\sigma²})(z-t)dt+\\
 & \int_{z}^{\infty}\exp(-\frac{(t-z_{i})²}{2\sigma²})(t-z)dt-\int_{z}^{\infty}\exp(-\frac{(t+z_{i})²}{2\sigma²})(t-z)dt\:)\:,\end{array}$\\
and finally the potential is:\\
$\begin{array}{ll}
V(z)= & -2\sqrt{2\pi}\sigma\, q_{i}\,(\exp(-\frac{(z-z_{i})²}{2\sigma²})-\exp(-\frac{(z+z_{i})²}{2\sigma²}))\\
 & -(z-z_{i})\,2\pi\, q_{i}\,(Erf[\frac{(z-z_{i})}{\sqrt{2}\sigma}]-Erf[\frac{(z+z_{i})}{\sqrt{2}\sigma}])\,.\end{array}$\\
From this function, the electric field across a single slab, centered
at $z=0$, is:

$E(z)=-\frac{\partial V(z)}{\partial z}=2\pi\, q_{i}\,(\, Erf[\frac{(z-z_{i})}{\sqrt{2}\sigma}]-Erf[\frac{(z+z_{i})}{\sqrt{2}\sigma}]\,)$.

When $\sigma\rightarrow0$, the functions tend to those found in electrostatics
textbooks.

\subsection*{B2: Several neutral slabs model for the quasi-2D macroscopic electric
field:}

The typical charge distribution of our samples of amphiphilic bilayers
can be decomposed in a linear superposition of neutral slabs. Replacing
for each slab the uniformly charged faces by gaussian distributions
of charges along \emph{z}, of total charge $\pm q_{i}$ each one,
width $\sigma_{i}$ and located at $\pm z_{i}$, the preceding formulae
in C.1 can be easily extended to any number of slabs and the final
results are:\\

$\rho(z)=\sum_{i}\,\frac{q_{i}}{\sqrt{2\pi}\sigma_{i}}\exp(-\frac{(z-z_{i})²}{2\sigma²_{i}}))\quad,$

$V(z)=-2\sqrt{2\pi}\sum_{i}\,\sigma_{i}\, q_{i}(\exp(-\frac{(z-z_{i})²}{2\sigma_{i}²})-(\frac{z-z_{i}}{\sqrt{2}\sigma_{i}})\, Erf[\frac{(z-z_{i})}{\sqrt{2}\sigma_{i}}])\,.$

$E(z)=-\frac{\partial V(z)}{\partial z}=2\pi\sum_{i}\, q_{i}\, Erf[\frac{(z-z_{i})}{\sqrt{2}\sigma}].$

The accumulated energy of this superposition of slabs is:

$W=\frac{1}{2}$$\int_{-\infty}^{\infty}\,\rho(z)V(z)dz=-\sum_{i}\sum_{j}\, q_{i}q_{j}\frac{\sigma_{j}}{\sigma_{i}}(I_{1}+\sqrt{\pi}I_{2}).$

The integral $I_{1}$ can be solved using the well known property
of gaussians: the product of two gaussians is a third gaussian, in
this case:

$I_{1}=\int_{-\infty}^{\infty}\,\exp(-\frac{(z-z_{i})²}{2\sigma_{i}²})\exp(-\frac{(z-z_{J})²}{2\sigma_{j}²})dz=\frac{\sqrt{\pi}}{\sqrt{\gamma}}\exp(-\frac{(z_{i}-z_{j})²}{2(\sigma_{i}²+\sigma_{j}²)})$;
with $\gamma=\frac{(\sigma_{i}²+\sigma_{j}²)}{\sigma_{i}²\sigma_{j}²}$.\\

The second integral is a new mathematical solution and was solved
with the aid of refs. \cite{errorfunc-int,math}. From the first one
\cite{errorfunc-int} we obtain:

$I_{2a}=\int_{-\infty}^{\infty}\,\exp(-a(t-c)²)erf(t)dt=\sqrt{\frac{\pi}{a}}erf(\frac{\sqrt{a}c}{\sqrt{1+a}})$,
and from the second reference:

$I_{2b}=\int_{-\infty}^{\infty}\, t\,\exp(-a(t-c)²)erf(t)dt=c\, I_{2a}+\frac{1}{a\sqrt{1+a}}\exp(-ac²+\frac{a²c²}{1+a})$.
Finally, and taking

$a=(\sigma_{j}/\sigma_{i})²$ and $c=\frac{(z_{i}-z_{j})}{\sqrt{2}\sigma_{j}}$,
we calculate:

$I_{2}=\int_{-\infty}^{\infty}\,\exp(-\frac{(z-z_{i})²}{2\sigma_{i}²})(\frac{z-z_{j}}{\sqrt{2}\sigma_{j}})\, Erf[\frac{(z-z_{j})}{\sqrt{2}\sigma_{j}}])=\sqrt{2}\sigma_{j}I_{2b}$.

This formulae is extremely simple to include, and fast to calculate
in a MD or MC simulation. The application to other geometries is also
straightforward. These relationships are also valid to apply when
working with systems where there is a finite difference of voltage
through the slab, keeping in mind that the total charge of the slab
is zero.

\end{document}